\documentclass[12pt]{article}

\input{epsf}
\usepackage[nosort]{cite}
\usepackage[dvips]{graphicx}
\usepackage{amsmath}
\usepackage{epsfig}
\usepackage{amsfonts}
\usepackage{amssymb}
\usepackage{multirow}
\usepackage{array}

\usepackage{epsfig}
\usepackage{amsmath}
\usepackage{amssymb}

\newcommand{\be}{\begin{equation}}
\newcommand{\ee}{\end{equation}}
\newcommand{\bee}{\begin{equation*}}
\newcommand{\eee}{\end{equation*}}
\newcommand{\bea}{\begin{eqnarray}}
\newcommand{\eea}{\end{eqnarray}}
\newcommand{\bean}{\begin{eqnarray*}}
\newcommand{\eean}{\end{eqnarray*}}

\newcommand{\nn}{\nonumber}

\newcommand{\lp}{\left(}
\newcommand{\rp}{\right)}

\begin{document}

\setcounter{page}{0}
\thispagestyle{empty}

\begin{flushright}
CERN-PH-TH-2010-273, \,
SACLAY-T10-176
\end{flushright}

\vskip 8pt

\begin{center}
{\bf \LARGE { Hydrodynamic obstruction to bubble expansion }}
\end{center}

\vskip 12pt

\begin{center}
{\bf Thomas Konstandin$^{a}$ and \bf   Jos\'e M. No$^{b}$ }
\end{center}

\vskip 20pt

\begin{center}

\centerline{$^{a}${\it CERN Physics Department, Theory Division, CH-1211 
Geneva 23, Switzerland}}
\centerline{$^{b}${\it Institut de Physique Th\'eorique, CEA/Saclay, F-91191 
Gif-sur-Yvette C\'edex, France}}
\vskip .3cm
\centerline{\tt tkonstan@cern.ch,}
\centerline{\tt jose-miguel.no@cea.fr}
\end{center}

\vskip 13pt

\begin{abstract}
We discuss a hydrodynamic obstruction to bubble wall acceleration
during a cosmological first-order phase transition. The obstruction
results from the heating of the plasma in the compression wave in
front of the phase transition boundary. We provide a simple criterion
for the occurrence of the obstruction at subsonic bubble wall velocity
in terms of the critical temperature, the phase transition
temperature, and the latent heat of the model under consideration. The
criterion serves as a sufficient condition of subsonic bubble wall
velocities as required by electroweak baryogenesis.
\end{abstract}

\newpage


\section{Introduction}

Cosmological first-order phase transitions can lead to many
interesting phenomena during the evolution of the early universe, such
as electroweak baryogenesis \cite{Cohen:1990py} or the production of a
stochastic background of gravitational waves
\cite{Witten:1984rs,Kosowsky:1991ua,Kamionkowski:1993fg}. 
A first-order phase transition proceeds by bubble nucleation and
subsequent bubble expansion and one essential quantity in the
description of phenomena linked to the phase transition is the
velocity of the expanding bubble walls $\xi_w$. For example, the
electroweak baryogenesis mechanism is based on the diffusion of
particle asymmetries into the plasma in front of the bubble wall and
subsonic bubble walls are necessary to build up a baryon asymmetry. On
the other hand, fast moving walls are essential for the production of
a sizable amount of gravitational radiation by bubble
collisions~\cite{Kosowsky:1991ua, Kamionkowski:1993fg, Caprini:2007xq,
Huber:2008hg, Caprini:2009fx},
turbulence~\cite{Kosowsky:2001xp,Caprini:2006jb} or magnetic
fields~\cite{Caprini:2006jb}.

The analysis of the bubble wall velocity generally assume that after a
short period of acceleration of order $\sim 1/T$, (where $T$
represents the typical energy scale associated with the temperature or
latent heat of the transition) the pressure difference that drives the
bubble expansion is balanced by friction and the bubbles subsequently
expand with a constant speed. To quantify this friction requires
solving a coupled system of Boltzmann equations for all particle
species with a sizable coupling to the Higgs field. This intricate
calculation has so far only been performed in the Standard Model
(SM)~\cite{Moore:1995si} and in the Minimal Supersymmetric Standard
Model (MSSM)~\cite{John:2000zq} under the assumption of small wall
velocities.

On the other hand, in the limit of highly relativistic wall velocities
it is found that the friction in the plasma tends to a
constant~\cite{BM} (up to possible $\log(\gamma_w) $ corrections),
opening the possibility of continuously accelerating (runaway) bubble
walls when the pressure difference along the phase boundary
overcomes this threshold. This runaway behavior is realistic in many
models, under the assumption that no hydrodynamic obstruction
prohibits that the highly relativistic regime is reached.

The goal of this paper is to analyze one of these possible
obstructions based on the heating of the plasma in front of the phase
boundary during bubble expansion. Implicitly, this effect was already
observed in refs.~\cite{Moore:1995si, KurkiSuonio:1996rk, EKNS} where
finite wall velocities have been reported in the limit of vanishing or
at least very small friction. In the recent work~\cite{EKNS} this
result was obtained under the assumption that the temperature in the
Higgs wall is identified with the temperature in front of the phase
boundary. Furthermore, the analysis focused on models with an equation
of state similar to the standard model.  Here, we relax those two
assumptions and present a simple criterion for the occurrence of the
obstruction. If our criterion holds in a specific model, the wall
velocity is subsonic and electroweak baryogenesis is in principle
possible. The heating effect only provides an upper limit and a
concrete determination of the wall velocity still requires some
knowledge of friction. However, as long as friction is not too strong
(as it is e.g.~the case in the SM) the resulting wall velocities are
fairly accurate. Besides, the baryon asymmetry in electroweak
baryogenesis is not very sensitive to the wall velocity as long as it
is significantly below the speed of sound and large enough to avoid a
saturation of the sphaleron process (see e.g.~\cite{Carena:2000id}
or~\cite{Huber:2001xf}). Hence, the knowledge of the precise wall
velocity is not so relevant for baryogenesis as long as it is
subsonic.

The paper is organized as follows: in the next section the
hydrodynamic treatment of the system is reviewed, and we briefly
sketch the origin of the obstruction. In sec.~\ref{subsec_EoS} the
heating effect is discussed analytically in a system with a bag
equation of state. We then study how the heating effect impacts the
acceleration of the wall in a toy model and present a simple criterion
for the occurrence of the obstruction in sec.~\ref{sec_toy}.  In
sec. \ref{sec_friction} we briefly comment on the interplay between
the hydrodynamic obstruction and friction before we apply our results
to specific models and conclude in sec.~\ref{sec:conclusion}.

\section{Origin of the obstruction\label{sec_hydro}}

In this section we introduce the basic concepts and set up the
notation used for the hydrodynamic analysis of the system of expanding
Higgs bubbles\cite{Landau, Steinhardt:1981ct, bubble_growth}. For most
parts we use the notation of~\cite{EKNS}.

The dynamics of the combined ``wall-plasma" system is determined by
the equations of motion of the plasma and of the Higgs field. However,
in the present case it is advantageous to replace the equation of
motion of the plasma (which would be of Boltzmann type) by the
assumption of local thermal equilibrium and energy-momentum
conservation, leading to the hydrodynamic approximation. The
energy-momentum tensor of the Higgs field $\phi $ is given by
\be
\label{eq:TmunuHiggs}
T_{\mu\nu}^{\phi} = \partial_{\mu}\phi \partial_{\nu}\phi -g_{\mu\nu} 
\left[ \frac{1}{2} \partial_{\rho}\phi \partial^{\rho}\phi - V_0 (\phi)\right],
\ee
where $V_0 (\phi) $ is the renormalized vacuum potential. If the
plasma is locally in equilibrium its energy-momentum tensor can be parametrized as
\be
\label{eq:Tmunu}
T_{\mu\nu}^{plasma} = w \, u_\mu u_\nu  - g_{\mu\nu} \, p \ , 
\ee
where $w$ and $p$ are the plasma enthalpy and pressure, respectively.
The quantity $u_\mu$ is the four-velocity field of the plasma, related to 
the
three-velocity $\mathbf{v}$ by
\be
u_\mu = \frac{(1, \mathbf{v} )}{\sqrt{1-\mathbf{v}^2}}
= (\gamma, \gamma\mathbf{v} ) \ .
\ee
A constant $\phi$ background contributes to the total pressure [see
eq.~(1)] and from now on we will use $p$ for this total pressure,
including such contribution. The enthalpy $w$, the entropy density
$\sigma$ and the energy density $e$ are defined by
\be
w \equiv T\frac{\partial p}{\partial T}\ ,\quad
\sigma \equiv \frac{\partial p}{\partial T}\ ,\quad
e\equiv T\frac{\partial p}{\partial T} -p\ ,
\ee
where $T$ is the temperature of the plasma. One then has 
\be
w = e + p\ .
\ee

Conservation of energy-momentum is given by 
\be
\label{eq:T_cons}
\partial^\mu T_{\mu\nu} = \partial^\mu T_{\mu\nu}^{\phi} 
+ \partial^\mu T_{\mu\nu}^{\mathrm{plasma}} = 0 \ .
\ee
We are interested in a system where the bubble expands at a constant speed and,
assuming there is no time-dependence, eq.~(\ref{eq:T_cons}) reads in
the wall frame (with the wall and fluid velocities aligned in the $z$
direction)
\be
\partial_z T^{zz} = \partial_z T^{z0} = 0\ .
\ee
Integrating these equations across the phase boundary and
denoting the phases in front and behind the wall by subscripts $+$
(symmetric phase) and $-$ (broken phase) one obtains the matching
equations (in the wall frame):
\be
\label{eq:wall_constr}
w_+ v^2_+ \gamma^2_+ + p_+  = w_- v^2_- \gamma^2_- + p_-\ ,
\quad 
w_+ v_+ \gamma^2_+ = w_- v_- \gamma^2_-\ .  
\ee
From these equations we can obtain the
relations~\cite{Steinhardt:1981ct}
\be
\label{eq:vvs0}
v_+ v_- = \frac{p_+  - p_-}{e_+ - e_-}\ , \quad
\frac{v_+}{ v_-} = \frac{e_- + p_+ }{e_+ + p_-}\ . 
\ee
In a concrete model the thermodynamic potentials can be calculated in
the two phases and the temperature at which the phase transition
happens can be determined using the standard
techniques~\cite{cite:fate}. Still, there are three unknown quantities
($T_-$, $v_+$ and $v_-$) and only two equations (\ref{eq:vvs0}), so
that up to this point all hydrodynamically viable solutions are
parametrized by one parameter. We will parametrize the solution by its
wall velocity $\xi_w$.

Ultimately, the wall velocity is deduced from the equation of motion
of the Higgs
\be
\label{HEOMprior}
\square \phi + \frac{\partial {V_0}}{\partial \phi} 
+\sum_i \frac{dm_i^2}{d\phi} \int \frac{d^3p}{(2\pi)^32E_i} f_i(p)= 0 \ ,
\ee
where $f_i$ denotes the particle distribution function of the $i$th
species. By decomposing
\be
 f_i(p)=f_i^{eq}(p)+\delta f_i(p)\ ,
\ee
where $f_i^{eq}=1/[\exp{(E_i/T)}\mp 1]$ is the equilibrium distribution 
function with $E_i^2=p^2+m_i^2$, eq.~(\ref{HEOMprior}) takes 
the simple form   
\be
\label{HEOM}
\square \phi + \frac{\partial {\cal F}}{\partial \phi} 
- {\cal K}(\phi) = 0 \ .
\ee
The second term contains the free energy $\cal F$ that drives the
expansion of the bubble and ${\cal K}(\phi)$ stands for the friction
term that arises from deviations of the particle distributions in the
plasma from equilibrium. In principle, calculation of ${\cal K}(\phi)$
involves solving a coupled system of Boltzmann equations for all
particle species with a sizable coupling to the Higgs field. This
intricate calculation has been performed in the Standard
Model~\cite{Moore:1995si} and in the MSSM~\cite{John:2000zq} and under
the assumption that the deviation from thermal equilibrium is small,
i.e. $ \delta f_i(p)\ll f_i(p)$, which is only true for weakly
first-order phase transitions.

The aim of the present work is to show that the bubble wall velocity
can approach a subsonic value even if the friction is very small -
contrary to the naive expectation of supersonic velocities or even
runaway behavior in the limit of very small friction. The occurrence
of this effect can be understood by inspecting
eq.~(\ref{HEOM}). Assuming that in the steady state the bubble is
large enough so that one can use the planar limit, one obtains by
integration the pressure (in the wall frame) that drives the expansion
\be
\label{eq:MS_EoM}
F_{dr} = \int \, dz \, \partial_z \phi\ \frac{\partial {\cal F}}{\partial \phi}
=\int \, dz \, \partial_z \phi \, {\cal K} = \, F_{fr} \ .
\ee
The physical interpretation of this equation is that the change of
pressure in the wall drives the expansion of the bubble and this
driving force $F_{dr} $ is ultimately balanced by the friction force
$F_{fr} $ in order to reach a stationary state. Without the influence
of the bubble, this change in pressure is always positive, since
nucleation requires the temperature in the symmetric phase to be below
the critical one. However, some particles are reflected at the phase
boundary, leading to a heating effect in front of the wall. Hence the
temperature experienced by the phase boundary is increased and a
hydrodynamic obstruction can occur when the bubbles accelerate while
building up a compression wave in front of the Higgs wall. At some
velocity, the average temperature\footnote{We will be more specific
what the average temperature is in sec.~\ref{sec_toy}.} in the wall
might approach the critical one, the driving force goes to zero and
the bubble cannot further accelerate even in the limit of vanishing
friction as defined by (\ref{HEOM}) and (\ref{eq:MS_EoM}).

As a final remark in this section, notice that there is a certain
freedom in splitting the last term of (\ref{HEOMprior}) into
equilibrium and friction contributions. At the critical temperature
$T_c$ (where the pressure in the two phases is equal), the only
solution to (\ref{HEOM}) is a static wall. The plasma is at rest and
the temperatures on both sides of the Higgs wall coincide: the system
is in equilibrium everywhere. For a temperature below the critical
one, all deviations from equilibrium in the particle distributions are
proportional to the wall velocity and also the friction term vanishes in
the limit of zero velocity. This leaves a certain arbitrariness in the
temperature profile of the equilibrium distributions $f^{eq}_i$ in
(11) along the wall and different choices for the temperature profile
in the equilibrium distributions will result in different friction
terms.

A necessary feature of the temperature profile is that it interpolates
between the two values $T_+$ and $T_-$. This ensures that the
deviations from equilibrium $\delta f_i$ vanish away from the bubble
wall. Besides, an upper bound on the wall velocity will be derived in
the following by neglecting the friction term. Hence, one has to
ensure that the friction term is always positive and reduces the wall
velocity for the chosen temperature profile.  A very convenient choice
for the temperature profile is obtained by enforcing conservation of
the two energy-momentum tensors resulting from the particle
distributions $f^{eq}_i$ and $\delta f_i$ separately. First, this has
the advantage that in the limit of infinite interaction rates the
particle distribution functions $f^{eq}_i$ are the physical ones and
$\delta f_i$ will vanish. Second, for this choice entropy is conserved
when the deviations $\delta f_i$ are neglected. A simple
phenomenological approach for the friction term shows (see
e.g.~ref.~\cite{KurkiSuonio:1995pp}) that negative friction terms
would lead to entropy decrease. This indicates that in general the
H-theorem ensures that the expansion around this equilibrium leads to
positive friction terms that reduce the wall velocity. We will
essentially utilize this choice for the temperature profile in
section~\ref{sec_toy}.

\section{The heating effect\label{subsec_EoS}}

In this section we discuss a system whose equation of state is in the
broken phase given by 
\be
\label{eosm}
p_- = \frac{1}{3}a_- T_-^4 + \epsilon \ ,\quad e_- = a_- T_-^4
- \epsilon\ ,
\ee
where $\epsilon$ denotes the false-vacuum energy resulting from the
Higgs potential, while in the symmetric phase
\be
\label{eosp}
p_+ = \frac{1}{3}a_+ T_+^4\ , \quad
e_+ = a_+ T_+^4 \ ,
\ee
with a different number of light degrees of freedom across the wall
and therefore different values $a_+$ and $a_-$ (with $a_+>a_-$) and
different temperatures on both sides of the wall. These expressions
correspond to the so-called bag equation of state. This approximation
works reasonably well when the Higgs vev does not change much in
between the critical temperature and zero temperature and particles
can be independently of the temperature divided into ``light'' and
``heavy'' in the two phases. In particular, this is the case for
models where the potential barrier between the broken and symmetric
phases is present even at vanishing temperature, generally leading to
relatively strong phase transitions.

For temperatures close to the critical one, it is more physical to
parametrize the pressure and energy density in the symmetric and
broken phases by
\be
p_- \simeq \frac13 a_+ T_-^4 - \ell_c (T_-/T_c - 1) \ , \quad
p_+ \simeq \frac13 a_+ T_+^4 \ , 
\ee
\be
e_- \simeq a_+ T_-^4 - \ell_c \ , \quad
e_+ \simeq a_+ T_+^4 \ , 
\ee
and $\ell_c$ is the latent heat at the critical temperature.
Comparing with the bag equation of state one can identify
\be
\label{eq:def_latheat}
\ell_c = \frac43 (a_+ - a_-) T_c^4 = 4 \epsilon \ . 
\ee
Our analysis will be done mostly assuming the bag equation of state in
order to facilitate comparison with the existing literature. However,
when rephrased in terms of the latent heat all results are equally
applicable to models with a weak phase transition where the bag
equation of state does not hold.

Using the bag equations of state (\ref{eosm}) and (\ref{eosp}) in 
eq.~(\ref{eq:vvs0}) we get
\bea
\label{eq:vvs}
v_+ v_- &=& \frac{1 - (1-3\alpha_+) r }
{3 - 3( 1 + \alpha_+ )r} \ , \nn \\ 
\frac{v_+}{ v_-} &=& \frac{3  + (1-3\alpha_+) r}
{1 + 3(1 + \alpha_+ )r} \ , 
\eea
where we defined 
\be
\alpha_+\equiv\frac{\epsilon}{a_+ T_+^4} = \frac{l_c}{4 a_+ T_+^4}\ ,\quad  
r\equiv\frac{a_+T_+^4}{a_-T_-^4}\ .
\ee 
The quantity $\alpha_+$ is the ratio of the vacuum energy to the radiation 
energy density and typically characterizes the ``strength" of the phase 
transition: the larger $\alpha_+$ the stronger the phase transition.
These two equations can be combined to give
\be
\label{eq:vvs2}
v_+ = \frac{1}{1+\alpha_+}\left[ \left(\frac{v_-}{2}+\frac{1}{6
v_-}\right) \pm \sqrt{\left(\frac{v_-}{2}+\frac{1}{6 v_-}\right)^2 +
\alpha_+^2 +\frac{2}{3}\alpha_+ - \frac{1}{3}} \right],
\ee
so that there are two branches of solutions, corresponding to the
$\pm $ signs in eq. (\ref{eq:vvs2}). 

Generally, there are three different types of solutions for the bulk
fluid motion (recent reviews on the topic can be found in~\cite{EKNS}
and \cite{Leitao:2010yw}): Detonations, deflagrations and hybrid
solutions. In detonations, the wall expands at supersonic velocities
and the vacuum energy of the Higgs leads to a rarefaction wave behind
the bubble wall, while the plasma in front of the wall is at rest. In
this case, the wall velocity is $\xi_w = v_+ > v_-$, and therefore
detonations are identified with the $+$ branch of solutions in
(\ref{eq:vvs2}). In deflagrations, the plasma is mostly affected by
reflection of particles at the bubble wall and a compression wave
builds up in front of the wall while the plasma behind the wall is at
rest. In this case, the wall velocity is identified with $\xi_w = v_-
> v_+$, corresponding to the $-$ branch of solutions in
(\ref{eq:vvs2}). ``Pure" deflagrations are subsonic, while the hybrid
case occurs for supersonic deflagrations where both effects
(compression and rarefaction wave) are present. In the following we
are concerned with deflagrations.

In the following we present a rough constraint for the occurrence
of the obstruction. We assume that the bag equation of state holds
even locally in the wall such that the pressure is given by the simple
expression $p = \frac13 a(\phi) T^4$. Then, the driving force
(\ref{eq:MS_EoM}) reads
\be
\label{eq:force}
F_{dr} = 
\int \, dz \, \partial_z \phi\ \frac{\partial {\cal F}}{\partial \phi} =
\epsilon - \frac13 \int \, da \,T^4\ .
\ee
For a monotonously changing temperature in the wall, the last
contribution is bounded by (we will see later on that $T_-<T_+$)
\be
\label{eq:Fdr_bounds}
\frac13 (a_+ - a_-) T_-^4 
< \epsilon - F_{dr} <
\frac13 (a_+ - a_-) T_+^4 \ .
\ee
Analogously to $\alpha_+$ one can define
the quantities $\alpha_N$, $\alpha_-$ and $\alpha_c$ as the vacuum
energy normalized to the energy density in the symmetric phase, behind
the wall and at the critical temperature
\be
\alpha_- = \frac{\epsilon}{a_- T_-^4} \ , \quad
\alpha_N = \frac{\epsilon}{a_+ T_N^4} \ , 
\ee
\be
\alpha_c = \frac{\epsilon}{a_+ T_c^4} = 
\frac13 \frac{a_+ - a_-}{a_+} < \frac13 \ .
\ee
The nucleation of bubbles happens at temperatures below the critical
one, $\alpha_c < \alpha_N$. After nucleation, the bubble wall
accelerates, increasing $T_+$ and therefore decreasing $\alpha_+$. Likewise,
$T_-$ is expected to increase and $\alpha_-$ to decrease during the
acceleration. If the phase transition is sufficiently weak, this
heating effect eventually leads to a vanishing driving force and
the acceleration of the wall ceases. According to
(\ref{eq:Fdr_bounds}) this has to happen somewhere in the window
\be
\label{eq:ob_exist}
T_- < T_c < T_+ \ .
\ee
We will show that for small wall velocities the relation $T_+ > T_- >
T_N$ holds, so the hydrodynamic obstruction definitely occurs in
models where $T_N$ is very close to $T_c$. 

\begin{figure}[ht]
\begin{center}
\includegraphics[width=0.95\textwidth, clip ]{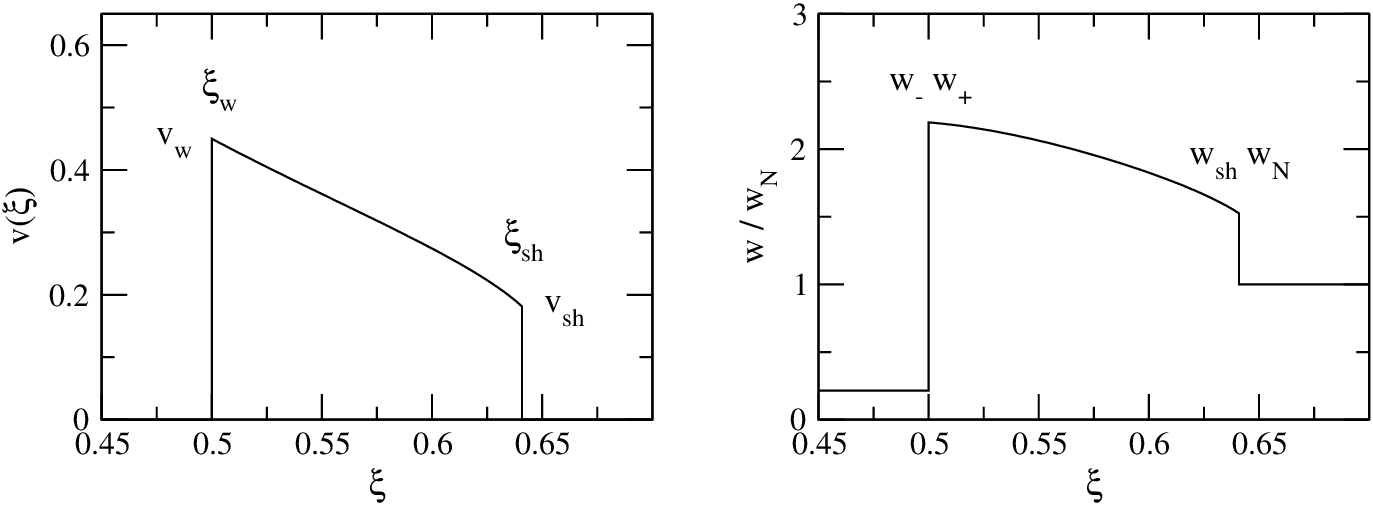}
\caption{\label{fig:deflag}
\small 
Example for a deflagration. The left plot shows the fluid velocity
$v(\xi)$ while the right plot shows the enthalpy $w(\xi)$.}
\end{center}
\end{figure}

In the following we briefly review the solutions of the plasma
velocity in the limit of small wall velocities $\xi_w$ as done
in~\cite{EKNS}. For deflagrations, the fluid in the bubble is at rest
and changes in the bubble wall to a finite value according to
(\ref{eq:vvs2}) where $v_-$ is identified with the wall velocity
$\xi_w$.  The fluid motion in the compression wave is then determined by
hydrodynamics. Since there is no intrinsic macroscopic length scale
present in the system, solutions to the hydrodynamic equations are
self-similar and only depend on the combination $\xi = r/t$, where $r$
denotes the radial coordinate of the bubble and $t$ the time since
nucleation. The plasma then fulfills the equations
\be
2 \frac{v}{\xi} = \gamma^2 (1 - v \xi) 
\left[ \frac{\mu^2}{c_s^2} -1 \right] \partial_\xi v,
\ee
and
\be
\label{eq:w_diff}
\frac{\partial_\xi w}{w} =4 \gamma^2 \, \mu(\xi,v) \partial_\xi v \ ,
\ee
where $c_s=1/\sqrt{3}$ denotes the velocity of sound in the plasma and
$\mu(\xi, v)$ is the Lorentz-transformed fluid velocity
\be
\mu(\xi, v) = \frac{\xi - v }{1 -\xi v} \ .
\ee
An example of the fluid motion and the enthalpy in case of a
deflagration is shown in Fig.~\ref{fig:deflag}.

The boundary conditions for deflagrations read in the rest frame
\be
v_w \equiv v (\xi_w) = \mu (v_-, v_+ ) \ ,
\ee
and the fluid velocity in the compression wave is for small wall and
fluid velocity ($\xi_w, v \ll 1$) given by
\be
v(\xi) \simeq v_w \frac{\xi^2_w}{\xi^2} \ , \quad 
\xi \in [\xi_w ,c_s] \ .
\ee
The enthalpy changes in the compression wave between the Higgs wall
and the plasma in front of the shock by a factor (there is also a jump
in the enthalpy in the shock front which is however of order
$\xi_w^3$)
\be
\frac{T^4_+}{T^4_N} =\frac{\alpha_N}{\alpha_+} = \frac{w_+}{w_N} \simeq \exp 
\left[ 4 \int_{v(c_s)}^{v(\xi_w)} \, \mu(\xi,v) \, dv \right] 
\approx  \exp \lp 8 \, \xi_w v_w \rp \ . 
\label{eq:apan}
\ee
So in this limit indeed $T_+ > T_N$, as required for an obstruction
(\ref{eq:ob_exist}). Using the bag equation of state and
(\ref{eq:vvs2}) one finds
\be
\label{eq:vvbound}
\frac{v_+}{v_-} \simeq 1 - 3 \alpha_+ -6 \alpha_+ \xi_w^2 \ ,
\ee
and hence 
\be
\label{eq:vw_small}
v_w \approx 3 \alpha_+ \xi_w \ . 
\ee
If the obstruction would occur as soon as $T_+$ exceeds the critical
temperature, the wall velocity would be
\be
\label{eq:estv1}
\xi_w^2 = \frac{\log \frac{T_c}{T_N}}{6 \alpha_c} \ .
\ee
Nevertheless, $T_-$ is typically smaller than $T_+$, so a
more conservative bound arises from the pressure in the broken phase, 
that is determined using
\be
\frac{w_+}{w_-} = 
\frac{v_- \gamma_-^2}{v_+ \gamma_+^2} 
\approx  \frac{1 + 12 \alpha_+ \xi_w^2}{1-3\alpha_+}, 
\ee
such that 
\be
\label{eq:bagTpTm}
\frac{T_+^4}{T_-^4} = \frac{a_-}{a_+} \frac{w_+}{w_-} 
= \frac{1 - 3 \alpha_c}{1 - 3 \alpha_+} 
\left( 1 + 12 \alpha_+ \xi_w^2 \right)  
\simeq 1 + 12 \alpha_+ \xi_w^2 \ ,
\ee
and hence (using (\ref{eq:apan}))
\be
\label{eq:estTmoTN}
\log\frac{T_-}{T_N} \simeq 3 \xi_w^2 \alpha_+ \ .
\ee
Assuming that the obstruction occurs for $T_-= T_c$ one finds
analogously to (\ref{eq:estv1}) (assuming $\alpha_c \simeq \alpha_+$ in a 
first approximation)
\be
\label{eq:estv2}
\xi_w^2 = \frac{\log \frac{T_c}{T_N}}
{3 \alpha_c} \ ,
\ee
and hence a slightly larger wall velocity. The obstruction should hence
be somewhere in the range 
\be
\label{eq:estv3}
\xi_w^2 \in  \lp \, 3 \pm 1 \, \rp \, \frac1{12} \,
\frac{\log \frac{T_c}{T_N}} {\alpha_c} \ .
\ee

However, even if this criterion leads to a supersonic wall velocity,
the obstruction might still be present, since for sizable wall
velocities the above approximation underestimates the effect of the
heating. In order to show this last statement, we assume a large but
subsonic wall velocity $\xi_w \lesssim c_s$. When the wall velocity
becomes supersonic, the expansion mode by deflagration is prone to
decay into a detonation and we assume that the obstruction is absent
in this case~\cite{KurkiSuonio:1996rk}. In the following we derive
some analytic results under the assumption $\alpha_+ \ll 1$ and then
present some numerical results for $\xi_w = c_s$ and arbitrary
$\alpha_+$. In this case we obtain for the plasma velocity in front of
the wall (using (\ref{eq:vvs2}))
\be
v_+ \simeq c_s \frac{1- \sqrt{ \alpha_+ (2 + 3 \alpha_+)}}{1 + \alpha_+}
+ O((\xi_w-c_s)^2) \ .
\ee
Notice that positive velocities imply that $\alpha_+ < 1/3$. If
$\alpha_+$ is larger than this value, deflagrations are not possible
and the bubble expansion proceeds by supersonic detonations. In the
wall frame the plasma has in front of the wall the velocity
\be
\label{eq:vw_large}
v_w = \mu(\xi_w, v_+) \simeq \frac32 \lp (\xi_w - c_s) + \sqrt{2 \alpha_+ /3} \rp \ ,
\ee
which is obviously only a valid approximation for $(c_s - \xi_w)
\lesssim \sqrt{2 \alpha_+/ 3}$. Comparing (\ref{eq:vw_small}) with
(\ref{eq:vw_large}) shows that for wall velocities close to the speed
of sound the heating effect in front of the wall is much more
efficient than in the limit of small wall velocities. The bound
(\ref{eq:estv3}) is hence a conservative one.

In the compression wave the plasma velocity decreases to a value
$v_{sh}$ and then drops to zero in the shock front. The decrease in
enthalpy in the compression wave is of order (using (\ref{eq:w_diff}))
\be
\log \frac{T_+}{T_{sh}} \simeq c_s (v_w - v_{sh}) 
+ O(v_w^2, \, v^2_{sh}) \ , 
\ee
whereas the jump in enthalpy in the shock front is of order (using
(\ref{eq:wall_constr}) in the shock front)
\be
\frac{T^4_{sh}}{T^4_N} \simeq \exp(4 c_s v_{sh} ) + O(v^3_{sh}) \ .
\ee
We therefore obtain
\be
\label{eq:ToTest1}
\log \frac{T_+}{T_N} \simeq c_s v_w  \simeq 
\frac32 c_s \lp (\xi_w - c_s) + \sqrt{2 \alpha_+/3} \rp \ ,
\ee
which is in leading order independent of the value of the fluid
velocity at the shock front $v_{sh}$ and the fluid profile in the
compression wave.

\begin{figure}[ht]
\begin{center}
\includegraphics[width=0.6\textwidth, clip ]{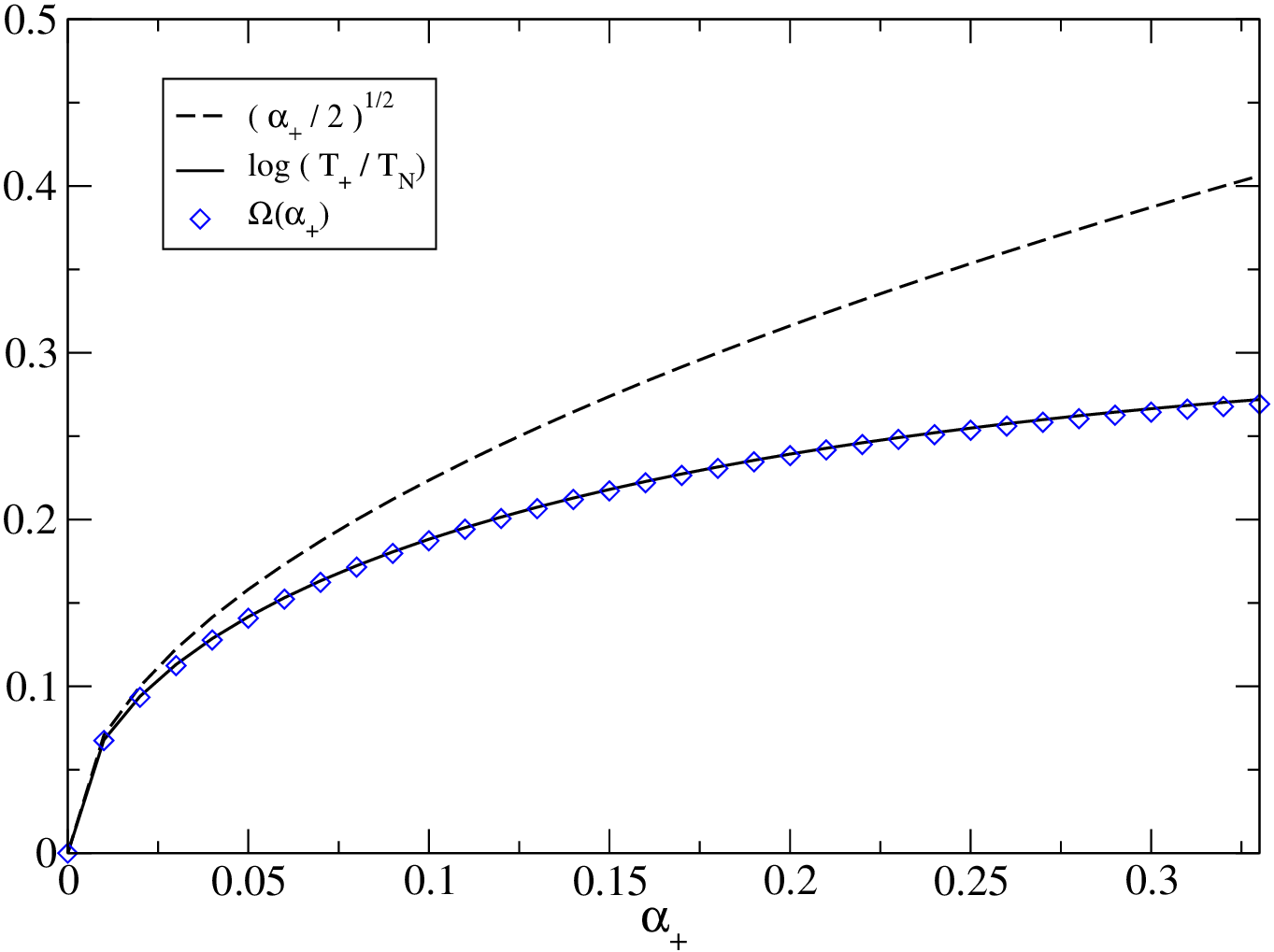}
\caption{\label{fig:omfit}
\small 
The heating effect assuming a wall velocity of the speed of sound,
$\xi_w=c_s$. The plot shows $\log T_+/T_N$ as a function
$\alpha_+$. The dashed line is the leading order result
(\ref{eq:ToT_leading}) while the straight line is the fit
(\ref{eq:ToT_fit}) to the numerical result.}
\end{center}
\end{figure}
To derive a simple criterion for the occurrence of the obstruction, we
assume again that the average temperature in the wall is close to
$T_+$ and that supersonic deflagrations can decay into detonations. For
a fixed value of $\alpha_+$, one can then determine from
(\ref{eq:ToTest1}) in leading order in $\alpha_+$
\be
\label{eq:ToT_leading}
\log \frac{T_+}{T_N}  \simeq \sqrt{\alpha_+/2} \ . \quad (\xi_w = c_s)
\ee
The full numerical result is shown in Fig.~\ref{fig:omfit}, and a
formidable fit is given by
\be
\label{eq:ToT_fit}
\log \frac{T_+}{T_N} \simeq \sqrt{\alpha_+/2} 
- \frac3{10} \alpha_+ - \frac15 \alpha_+^{3/2} \equiv \Omega(\alpha_+) \ .\quad (\xi_w = c_s)
\ee
If the obstruction would occur as soon as the temperature in front of
the wall $T_+$ surpasses the critical temperature $T_c$, an easy and
model independent criterion for the occurrence of the obstruction
would be
\be
\label{eq:crit_TpeqTc}
\log\frac{T_c}{T_N} < \Omega(\alpha_c) \ .
\ee
However, the enthalpy in the bubble interior is in leading order
(which is $O(\sqrt{\alpha})$) given by
\be
\label{eq:TmTn}
w_- \approx (1 - 4 c_s v_w) w_+ \approx w_N \ ,
\ee
and so the temperature in the bubble is close to the nucleation
temperature. In the limit $\alpha_c \to 1/3$, the temperature behind
the wall even vanishes. Hence, one expects that the temperature $T_+$
has to surpass the critical temperature $T_c$ significantly to obtain
an average temperature in the wall that is close to the critical
one. To obtain the wall velocity corresponding to the hydrodynamic
obstruction (if present) requires the knowledge of the temperature
profile in the wall in order to determine the average temperature
experienced by the Higgs. This is the topic of the next section.

\section{Local equations of motion in a toy model \label{sec_toy}}

In the last section, we presented results for the heating effect in
front of the wall during the bubble expansion. In particular, we
exemplified how to calculate the temperature on both sides of the
Higgs wall, $T_+$ and $T_-$, as functions of the wall velocity
$\xi_w$, the strength of the phase transition $\alpha_+$ and the
nucleation temperature $T_N$. The topic of the present section is to
connect this heating effect with the occurrence of the obstruction to
bubble acceleration. An exact treatment of this question would require
to determine the particle distribution functions by solving a system
of Boltzmann equations. The thermodynamic potentials in the wall could
then be obtained via (\ref{HEOMprior}). On the other hand, the
obstruction has to happen in the window given by
eq.~(\ref{eq:ob_exist}) and any simplifying assumption should lead to
reasonable bubble wall velocities. In this section, we assume that the
ansatz of local equilibrium (\ref{eq:Tmunu}) also applies in the
bubble wall. This approximation becomes exact in the limit that the
mean free path of the particles in the plasma is much shorter than the
bubble wall thickness. Even though the mean free path of some
particles is much larger than the wall thickness, we expect that the
free energy should not depend too much on the out-of-equilibrium
features of the particle distribution functions since energy-momentum
conservation generally holds (see also the comments at the end of
section~\ref{sec_hydro}).

\subsection{Solving the local equations of motion}

To obtain the steady-state profiles across the phase transition wall
of quantities like the velocity, temperature and Higgs field, we
integrate the following system of coupled differential equations in
the planar limit:
\bea
&&\partial^2_z\phi + \frac{\partial 
p }{\partial\phi} - {\cal K}=0\ ,\label{eq:loc1}\\
&&\partial_z[\omega \gamma^2 v]=0\ ,\label{eq:loc2}\\
&&\partial_z\left[\frac{1}{2}(\partial_z\phi)^2+\omega \gamma^2 v^2+p
\right]=0\ \label{eq:loc3} \ .
\eea
The first equation is the Higgs equation (\ref{HEOM}) and to be
specific we use the friction term (see~\cite{EKNS} and
also~\cite{Megevand:2009ut} for a similar approach)
\be
{\cal K} = -T_N\, \eta\, v\partial_z\phi \ .
\ee
The particular form of the friction term is not relevant since we are
interested in the regime of small friction, $\eta \to 0$.  The two
extra equations correspond to the differential (and static) form of
energy-momentum conservation (\ref{eq:T_cons}). They result from the
assumption that the system is locally in equilibrium, also on scales
that are comparable with the wall thickness. Their integration across
the wall gives immediately Steinhardt's matching conditions
(\ref{eq:wall_constr}).

In order to solve the system of equations, we proceed as follows. The
basis quantity is the free energy of the system ${\cal F}(\phi, T)$
from which all the thermodynamic potentials can be derived. We
determine the nucleation temperature of the system using the standard
techniques of the bounce solution~\cite{cite:fate}. With a specific
wall velocity $\xi_w$ and this information, the hydrodynamic system
(\ref{eq:vvs0}) can be solved. Finally, we solve the system of
differential equations (\ref{eq:loc1})-(\ref{eq:loc3}) and determine
which friction coefficient $\eta$ corresponds to the wall velocity
$\xi_w$ we have chosen. When the wall velocity is increased, the
friction coefficient eventually turns negative. This signals the
occurrence of the hydrodynamic obstruction. The numerical results
obtained by this procedure depend neither on the bag equation of state
nor on the assumption of small fluid or wall velocities.

The system of equations can be solved in the following way: First,
equation (\ref{eq:loc2}) can be used to obtain the fluid velocity as a
function of the Higgs vev and the temperature
\be
v (\phi, T) = -\frac{\omega(\phi,T)}{2 \omega_+ \gamma_+^2 v_+}
+ \sqrt{1 + \frac{\omega(\phi,T)^2}{4 \omega^2_+ \gamma_+^4 v^2_+}} \ .
\ee
Then eq.~(\ref{eq:loc3}) can be used to obtain a closed expression for $\partial_z
\phi(\phi, T)$
\be
\frac12 (\partial_z \phi(\phi, T))^2 = 
\omega \gamma^2 v^2 + p - \omega_+ \gamma_+^2 v_+^2 - p_+ \ .
\ee
Finally, the Higgs equation (\ref{eq:loc1}) can be recast using
(\ref{eq:loc3}) as
\be
\partial_z \phi \frac{\partial p}{\partial \phi}
- \partial_z \phi \, {\cal K} = \partial_z (\omega \gamma^2 v^2 + p) \ , 
\ee
which leads to 
\be
\label{eq:loc_dphidT}
\frac{d \phi}{d T} = 
\frac{\partial_T (\omega \gamma^2 v^2 + p)}
{{\cal K} - \partial_\phi (\omega \gamma^2 v^2)} \ .
\ee
For a fixed wall velocity $\xi_w$, one unique choice of $\eta$ leads
to the correct boundary conditions $\phi(T_\pm) = \phi_\pm$. The wall
velocity can then be varied to lead to the velocity corresponding to
the obstruction where $\eta=0$.

As a remark we note that solving the system of local equations
(\ref{eq:loc3}) also allows to determine the wall thickness. Usually
the wall thickness is determined for the nucleated bubbles from the
bounce solution. In principle the wall thickness during the bubble
expansion is expected to be smaller than this value since the
acceleration of the wall also compresses the wall. In practice we find
that this effect changes the wall thickness by only a few percent.

\subsection{A concrete (toy) model}

Consider the effective potential
\be
\label{eq:Vtoy}
V(\phi, T) = \lambda (\phi^2 - v^2)^2 - \frac{a_+}3 \, T^4 + c_1 T^2
\phi^2 - c_2 T \phi^3 \ .
\ee
This model is inspired by systems where the strength of the phase
transition comes from thermal contributions of bosons coupled to the
Higgs that add to the thermal potential terms proportional to $T
\phi^3$. This is for example the case in the light stop
scenario~\cite{Carena:1996wj} of the MSSM where the W-bosons and the
stops give rise to the cubic terms. Another example of this class of
models are hidden sector models~\cite{Espinosa:2008kw}. The parameters
$\lambda$, $c_1$ and $c_2$ can be exchanged for the Higgs mass $m_H$,
the critical temperature $T_c$ and the ratio $\phi_c / T_c$ according
to
\be
m^2_H = 8 \lambda v^2, \quad
\frac{\phi_c}{ T_c} = \frac{c_2}{2 \lambda}, \quad
T_c = \frac{4 \lambda v}{\sqrt{8 c_1 \lambda - 2 c_2^2}} \ .
\ee
The latent heat $\ell_c$ is given by 
\be
\ell_c \simeq T \, 
\partial_T \left[ V (T , \phi_c) - V (T , 0) \right]_{T=T_c} 
= 4 \lambda \phi_c^2 v^2 \ ,  
\ee
and hence 
\be
\label{eq:weak_alpha}
\alpha_c = \frac{\lambda \phi_c^2 v^2}{a_+ T_c^4} \ .
\ee

The nucleation temperature can be determined using the expression for
the tunneling action in the thin-wall approximation
\be
\label{eq:thin-wall}
S_3 / T = \sqrt{2}\frac{4 \pi}{81} 
\frac{\lambda^{3/2} \phi_c^{9}}{\ell_c^2 T_N} 
\left( 1- \frac{T}{T_c} \right)^{-2} \ ,
\ee
yielding for the nucleation temperature ($S_3 / T_N \approx 140$)
\be
\label{eq:toyTcTn}
\log\frac{T_c}{T_N} \simeq \left( 1 - \frac{T_N}{T_c} \right)
\simeq 0.04 \, 
\frac{\lambda^{3/4} \phi_c^4}{\ell_c } \sqrt{\frac{\phi_c}{T_c}} \ . 
\ee
Using this in (\ref{eq:estv3}) one obtains
\be
\label{eq:main}
\xi_w^2 \in (\, 0.7 \div 1.4 \,) \, 
\lp \frac{g_*}{100} \rp 
\left(\frac{v}{m_H} \right)^{5/2}
 \lp \frac{T_c}{v} \rp^4 
\sqrt{\frac{\phi_c}{T_c}} \ ,
\ee
where the number of degrees of freedom in the symmetric phase $g_*$
is given by $g_* = 30 \, a_+ / \pi^2$. Hence the obstruction velocity
depends in the regime of weak phase transitions only weakly on
$\phi_c$. Still, the thin-wall approximation (\ref{eq:thin-wall}) is
not very accurate and we use numerical results for the nucleation
temperature in the following.

\begin{figure}[ht]
\begin{center}
\includegraphics[width=0.6\textwidth, clip ]{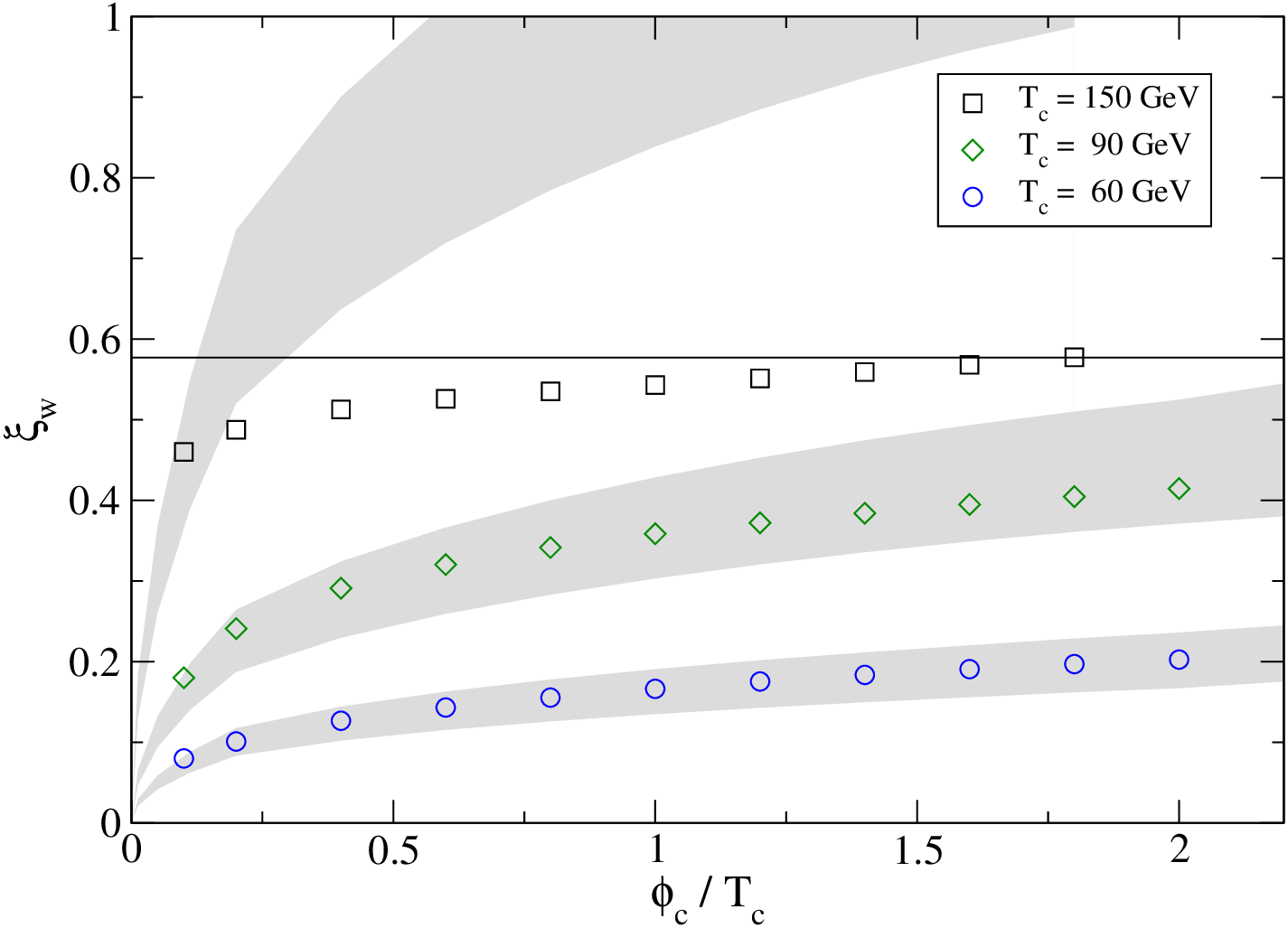}
\caption{\label{fig:toy}
\small 
The wall velocity of the obstruction versus $\phi_c/T_c$ for the
values $T_c=0.6$, $T_c=0.9$, and $T_c=1.5$. The Higgs mass is $m_H =
120$ GeV and the gray bands correspond to the estimate in~(\ref{eq:estv3}).  }
\end{center}
\end{figure}

The result from solving the local equations compared to the estimate
in (\ref{eq:estv3}) is shown in Fig.~\ref{fig:toy}. As expected, for
small velocities the estimate works rather well while it
underestimates the heating effect in front of the wall when the wall
velocity approaches the speed of sound.

In the last section we found the criterion (\ref{eq:crit_TpeqTc}) for
the occurrence of the obstruction under the assumption that the
temperature in front of the wall equals the average one. In the
following we discuss how this criterion is modified if this assumption
is replaced by the local system of equations. Still, one would like to
express the criterion for the occurrence of the obstruction in terms
of critical and nucleation temperature only, since these quantities
can be obtained by inspecting the effective potential without an
hydrodynamic analysis of the deflagration mode. Especially for weak
phase transitions one would expect that the criterion
(\ref{eq:crit_TpeqTc}) is only modified by a proportionality factor $c(\alpha_c)$
close to unity:

\be
\label{eq:crit_TpeqTc_2}
\log\frac{T_c}{T_N} < c(\alpha_c) \Omega(\alpha_c) \ .
\ee

\begin{figure}[ht]
\begin{center}
\includegraphics[width=0.6\textwidth, clip ]{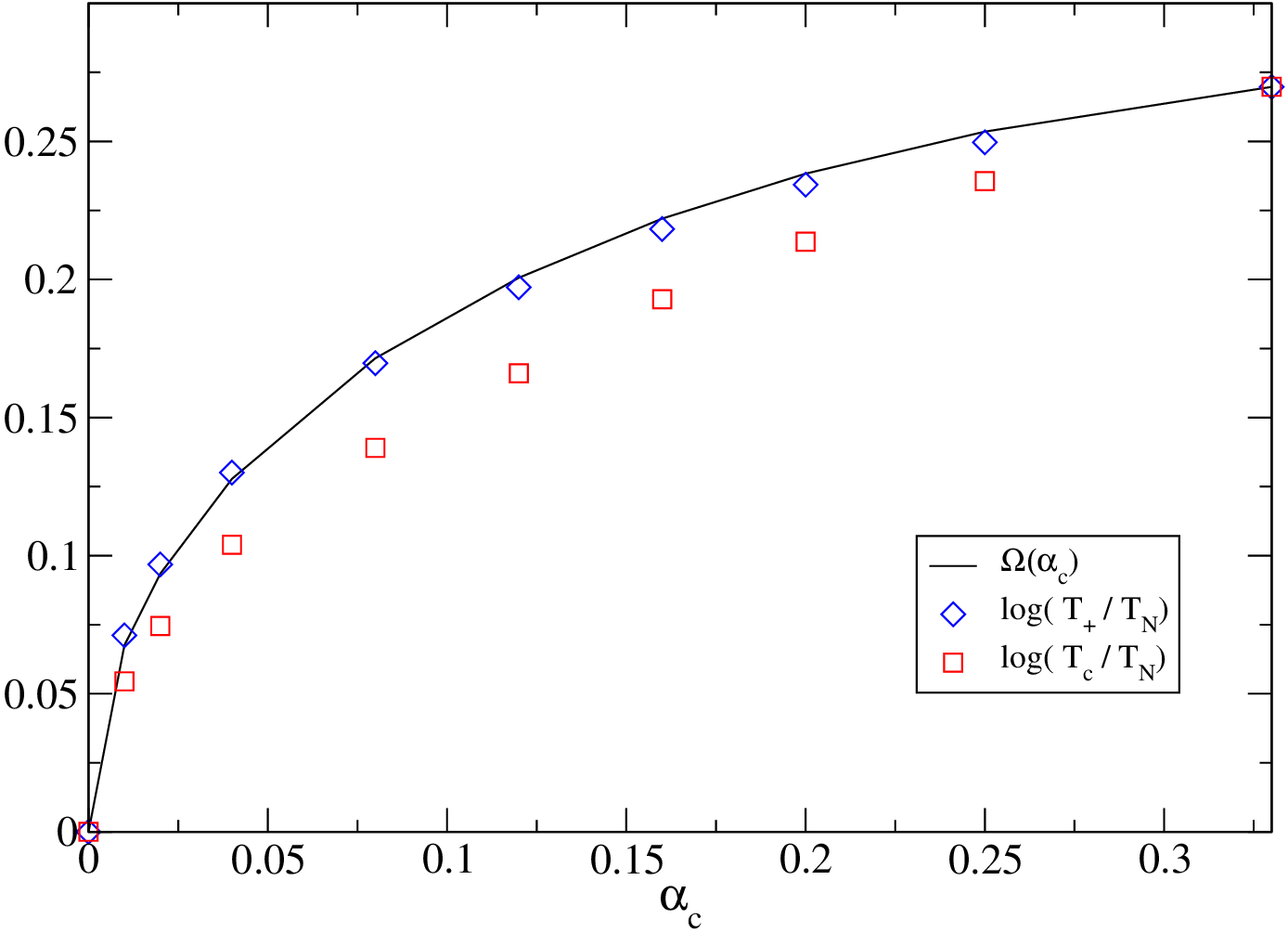}
\caption{\label{fig:strong}
\small 
The heating effect assuming a wall velocity of the speed of sound,
$\xi_w=c_s$. The plot displays the relation between the parameters of
the phase transition at critical temperature in relation to the
parameters in front of the phase boundary in the toy model
(\ref{eq:Vtoy}).  }
\end{center}
\end{figure}

In order to test this hypothesis, we determine the wall velocity of
the obstruction in the model (\ref{eq:Vtoy}). We keep $\alpha_c$ fixed
and vary $T_c$ in such a way that $\xi_w=c_s$. The functions
$\log\frac{T_+}{T_N}=\Omega(\alpha_+)$, $\log\frac{T_c}{T_N}$ and
$\Omega(\alpha_c)$ are plotted in the right panel of
Fig.~\ref{fig:strong}.  It turns out that the proportionality factor
in (\ref{eq:crit_TpeqTc_2}) is for small $\alpha_c$ close to
$\frac34$. This factor depends in principle on the model but it is
expected to be in the range $\left[\frac12, 1\right]$. The upper bound
results from the fact that the average temperature in the wall cannot
exceed $T_+$. The lower bound is due to the fact that the average
temperature cannot fall below $T_-$ and that for small $\alpha_c$ the
heating effect is behind the wall half as strong as in front (see
eqs.~(\ref{eq:estv1}) and (\ref{eq:estv2})).  Still, there is a
certain uncertainty in this factor\footnote{For a model with an
additional $\phi^6$ term in the Higgs potential the proportionality
factor also turns out to be close to $\frac34$ similar to the toy
model of weak phase transitions.}.

Interestingly, the criterion (\ref{eq:crit_TpeqTc}) becomes exact in
the limit $\alpha_c \to 1/3$.  This can be explained by noting that in
the limit $\alpha_c \to 1/3$, the velocity in front of the wall equals
the bubble wall velocity, $v_+=0$, and so the temperature behind the
wall $T_-$ has to drop to zero. Thus, the combination $\omega \gamma^2
v$ is constant in the wall and small. According to
(\ref{eq:loc_dphidT}) this implies
\be
\frac{d \phi}{d T} \simeq
- \frac{1}{\omega \gamma^2 v }
\frac{\partial_T p}{\partial_\phi v} \ ,
\ee
and therefore the Higgs vev changes quickly close to the broken phase
(where $v$ is small) and slowly close to the symmetric phase (where
$p$ is small). Hence in this limit one obtains $T_+ \to T_c$ and the
criterion (\ref{eq:crit_TpeqTc}) becomes exact. However, this behavior
could result from the assumption of local equilibrium and might not be
reproduced if the full set of Boltzmann equations are solved instead.

\section{Friction revisited\label{sec_friction}}

The analysis in the last few sections dealt with the hydrodynamic
obstruction in the limit of vanishing friction what gives an upper
limit on the wall velocity. In this section we indicate how both
effects - heating and friction - can be taken into account.

Our starting point is again the equation of motion of the Higgs field
(\ref{HEOM})
\be
\square \phi + \frac{\partial {\cal F}}{\partial \phi} 
- {\cal K}(\phi) = 0 \ .
\ee
where $ {\cal F}(\phi,T)$ denotes the free energy of the system and
${\cal K}(\phi) $ quantifies non-equilibrium effects that ultimately
lead to friction and hinder the bubble expansion. Integration of the
equation determines the acceleration rate of the bubble wall, which is
of the form
\be
\dot{\xi_w} \propto \left({\cal F}_--{\cal F}_+\right)(\bar T) - \xi_w \eta \ ,
\ee
where ${\cal F}_{\pm} $ are the free energies in the symmetric and
broken phase, and the coefficient $\eta $ parametrizes the
friction. The temperature $\bar T$ should be identified with the
average temperature in the wall, that can differ significantly from
$T_N $ due to the compression wave in front of the bubble wall in the
deflagration mode. Taking this effect into account leads (for small
wall velocities $\xi_w$) to an equation of the form
\be
\dot{\xi_w} \propto \left({\cal F}_--{\cal F}_+\right)(T_N) - \xi^2_w \kappa- \xi_w \eta \ ,
\ee
where $\kappa$ is (for small $\alpha_c$) approximately given by
\be
\kappa = \xi^{-2}_w \left(1-\frac{T_N}{\bar T} \right) 
\left. T_c \frac{\partial \left({\cal F}_--{\cal F}_+\right)}{\partial T}\right|_{T=T_c} 
\simeq \, \xi^{-2}_w \left(1-\frac{T_N}{\bar T} \right) \ell_c
\simeq 4 \alpha_c \ell_c \ .
\ee
This equation is only valid for weak phase transitions ($\alpha_c \ll
1$) and small wall velocities ($\xi_w \ll 1$) but a generalization can
be easily obtained by solving the hydrodynamic system along the lines
of sec.~\ref{subsec_EoS} in other regimes. For models where the
friction coefficient $\eta$ is known (in particular for models with a
particle content similar to the SM or MSSM), this approach can be used
to obtain accurate wall velocities.

\section{Discussion\label{sec:conclusion}}

We discussed the expansion velocity of nucleated bubbles in a
first-order phase transition. For the subsonic deflagration mode, the
expansion is not only hindered by friction (that results from
deviations from equilibrium) but also due to a heating effect by the
compression wave in front of the bubble wall. Even if friction is
neglected, the heating effect by itself can lead to an upper limit on
the expansion velocity of the bubbles.

In summary, a criterion for the occurrence of this obstructions at
subsonic wall velocities is given by
\be
\label{eq:crit_TpeqTc_final}
\log\frac{T_c}{T_N} < c(\alpha_c) \Omega(\alpha_c) \ ,
\ee
where $T_c$ and $T_N$ denote the critical and nucleation temperature
of the phase transition, respectively. The parameter $\alpha_c$
quantifies the strength of the phase transition and is given by
$\alpha_c = \frac{\ell_c}{4 e_c}$ where $\ell_c$ denotes the latent
heat and $e_c$ the energy density of the plasma in the symmetric
phase, both evaluated at the critical temperature. A fit to the
model-independent function $\Omega(\alpha_c)$ is given in
eq.~(\ref{eq:ToT_fit}). Finally, the function $c(\alpha_c)$ is a
model-dependent fudge factor that approaches unity for $\alpha_c \to
\frac13$ and a constant in the range $\left[\frac12, 1\right]$ in the
limit $\alpha_c \to 0$. For the model of weak phase transitions given
in (\ref{eq:Vtoy}) as well as the SM with an additional $\phi^6$ term
in the Higgs potential it is found to be close to $\frac34$ .

Let us discuss this result in the context of different
models. Consider the MSSM in the light stop
scenario~\cite{Carena:1996wj, Moreno:1998bq}. Using the
characteristics of the phase transition
\be
\phi_c \simeq T_c \simeq 95 \textrm{ GeV},\, 
m_H \simeq 120 \textrm{ GeV} \ ,
\ee
in (\ref{eq:weak_alpha}) and (\ref{eq:toyTcTn}) leads to
\be
\alpha_c = 2.6 \times 10^{-2}, \quad
\log \frac{T_c}{T_n} = 3.4 \times 10^{-3} \ ,
\ee
and an obstruction at the velocity $\xi_w
\simeq 0.18$. Hence, one can conclude even without the knowledge of the
friction coefficient that the wall velocity in the MSSM is subsonic
what is a necessary requirement for electroweak baryogenesis. An
analysis of friction without the heating effect~\cite{John:2000zq}
leads to a wall velocity $\xi_w \simeq (0.05 \div 0.1)$ what shows that
in this model friction is very relevant.

Next consider the SM with the (unrealistic) parameters
\be
\phi_c \simeq 60 \textrm{ GeV} \ ,\, 
T_c \simeq 120 \textrm{ GeV} \ ,\, 
m_H \simeq 90 \textrm{ GeV} \ ,
\ee
what corresponds to one parameter set in ref.~\cite{Moore:1995si} and yields
\be
\alpha_c = 2.3 \times 10^{-3} \ , \quad
\log \frac{T_c}{T_n} = 1.1 \times 10^{-3} \ ,
\ee
and hence $\xi_w \simeq 0.35$. In comparison, in \cite{Moore:1995si}
the heating effect as well as friction was taken into account and a
wall velocity in the range $\xi_w \in [0.35,\, 0.40]$ was found. The
result was not very sensitive towards changes in the interaction rates
such that the heating in front of the phase boundary is the dominant
effect hindering the wall expansion. This also fits well with the
results in \cite{Moore:1995ua} that neglected the heating effect and
found much larger wall velocities.

Let us discuss some cases where the wall velocity is still
unknown. Consider the nMSSM as discussed in~\cite{Apreda:2001us,
Huber:2006wf}. In this model, typical parameters for a phase
transition that is on the strong end of the spectrum
are\footnote{Notice that in~\cite{Apreda:2001us, Huber:2006wf}
the latent heat normalized to the energy density at
the nucleation temperature is denoted $\alpha$.}
\be
T_c \simeq 75 \textrm{ GeV} \ ,\, 
T_n \simeq 65 \textrm{ GeV} \ ,\, 
\alpha_c \simeq 8.9 \times 10^{-2} \ ,
\ee
what leads to a subsonic wall velocity according to
(\ref{eq:crit_TpeqTc_final}) and electroweak baryogenesis is feasible
in this model.

Finally, consider the SM enhanced by a large number of singlets from a
hidden sector as discussed in \cite{Espinosa:2008kw}. For a singlet
coupling to the Higgs of order $\zeta \simeq 1.0$ and a Higgs mass
$m_H = 125$ GeV one finds for example the following characteristics of
the phase transition\footnote{Also in~\cite{Espinosa:2008kw} the
latent heat normalized to the energy density at the nucleation
temperature is denoted~$\alpha$.}
\be
T_c \simeq 100 \textrm{ GeV} \ ,\, 
T_n \simeq 93 \textrm{ GeV} \ ,\, 
\alpha_c \simeq 1.9 \times 10^{-2} \ ,
\ee
and hence a wall velocity $\xi_w \simeq c_s$ [using the criterion for
large velocities (\ref{eq:crit_TpeqTc_final})]. For smaller values of
the scalar self-coupling $\zeta$ and/or larger Higgs masses the phase
transition becomes weaker such that for these parameters the wall
velocity is definitely subsonic. On the other hand, under the
assumption that the friction is similar to the SM, the wall velocity
is supersonic for larger values of $\zeta$ and/or smaller Higgs masses
and can even enter the runaway regime according to the analysis in
ref.~\cite{EKNS}.

\section*{Acknowledgment}

We are grateful to Geraldine Servant for helpful comments on the
manuscript. T.K.~acknowledges support from the European Commission the
European Research Council Starting Grant Cosmo@LHC.  The work of
J.M.N.~was partially supported by the European Community under the
contract PITN-GA-2009-237920 and by the Agence Nationale de la
Recherche.

\end{document}